\documentclass[draftcls, onecolumn,letterpaper,12pt]{ieeeconf}
\usepackage{mathrsfs}
\usepackage{epsfig}
\usepackage{amsmath,cite}
\usepackage{amsfonts, amssymb}
\usepackage{chemarrow}
\usepackage{graphicx}

\newcommand{\bpm}{\left(\begin{matrix}}
\newcommand{\epm}{\end{matrix}\right)}

\renewcommand{\baselinestretch}{1.1}

\begin{document}
\title{{\LARGE \textbf{Reconstruction of Arbitrary Biochemical Reaction
Networks: A Compressive Sensing Approach}}
}
\author{Wei Pan, ~Ye Yuan$^\dagger$ and ~Guy-Bart Stan \thanks{%
W.~Pan, Y.~Yuan and G.-B. ~Stan are with Centre for Synthetic Biology and
Innovation and the Department of Bioengineering, Imperial College London, United
Kingdom.}
\thanks{Y.~Yuan is also with Control Group, Department of Engineering, University of Cambridge, United Kingdom.}
\thanks{$^\dagger$For correspondence,
\tt\small  yy311@cam.ac.uk.}}
\date{15/2/2012}
\maketitle

\begin{abstract}
Reconstruction of biochemical reaction networks is a central topic in
systems biology which raises crucial theoretical challenges in system
identification. Nonlinear Ordinary Differential Equations (ODEs)
that involve polynomial and rational functions are typically used to model
biochemical reaction networks. Such nonlinear models make the problem of
determining the connectivity of biochemical networks from time-series
experimental data quite difficult. In this paper, we present a network
reconstruction algorithm that can deal with model descriptions under the
form of polynomial and rational functions. Rather than identifying the
parameters of linear or nonlinear ODEs characterised by pre-defined
equation structures, our methodology allows us to determine the nonlinear ODEs
structure together with their associated reaction constants. To solve the network
reconstruction problem, we cast it as a Compressive Sensing (CS) problem and use
Bayesian Sparse Learning (BSL) algorithms as an efficient way to obtain its solution.
\end{abstract}


\section{INTRODUCTION}\label{sec:introduction}

A long standing problem in systems biology is to reconstruct biochemical
reaction networks. Reconstruction means to identify both the topology and the
parameter values of some biochemical reaction networks. More specifically,
network reconstruction tries to recover the set of nonlinear ODEs associated with the
biochemical processes from time-series experimental data. A naive reconstruction
method consists in searching among all possible reactions the few that seem consistent
with the time series. The associated computational burden of such a naive approach is typically
horrendous even for network of modest dimensions.
Within the systems and control community, identification
of biochemical reaction networks, genetic regulatory networks in
particular, are quite active research areas \cite{control1,control2,control3,control4,control5,control6}.


Many linear and nonlinear functions can be used to describe the dynamics of
biochemical reaction networks in terms of biochemical kinetic laws, e.g.,
as first-order functions $f(\left[ S\right] )=\left[ S\right] ,$ mass action
functions $f(\left[ S_{1}\right] ,\left[ S_{2}\right] )=\left[ S_{1}\right]
\cdot \left[ S_{2}\right] ,$ Michaelis-Menten functions $f(\left[ S\right]
)=V_{\max }\left[ S\right] /(K_{M}+\left[ S\right] ),$ Hill functions $f(%
\left[ S\right] )=V_{\max }\left[ S\right] ^{n}/(K_{M}^{n}+[S]^{n}).$ Furthermore,
it is not uncommon that a single gene is regulated by more than one
transcription factor. In such situations, the combined effect of these regulators
on gene expression needs to be described by
a multi-dimensional input function. Such input functions typically take the form of ratio of polynomials involving the concentrations of the
input transcription factors $x_i,$ $i=1,\ldots ,n,$ for example, $%
f(x_{1},\ldots ,x_{n})=\sum\nolimits_i\beta
_i(x_i/K_i)^{n_i}/\left( 1+\sum\nolimits_i\beta
_i(x_i/K_i)^{m_i}\right) $, where $K_i$ is the activation or
repression coefficient for the transcription factor $x_i$, $\beta _i$
is its maximal contribution to gene expression, and the Hill coefficients are $%
n_i=m_i>0$ for activators and $n_i=0$, $m_i>0$ for repressors. These types of
functions have been shown to appropriately describe experimentally determined input
functions \cite{setty2003detailed}. More complicated expressions are also
possible, e.g., if the different transcription factors interact with each other at
the protein level \cite{buchler2003schemes}.

Our main objective is, given experimental time-series data, to identify both the interconnection topology (the form of the nonlinear functions) and their associated parameters.
Without prior knowledge on model structure, this is still an open problem in system identification.

During his plenary talk at the 50th IEEE CDC, Prof. Lenart Ljung emphasised on four opportunities for further research and development in system identification. Two of these were: (a) the use of sparse datasets and (b) the use of machine learning approaches \cite{ljung2011four}. The approach that we consider in this paper is aligned
with these two recommendations. Specifically, our approach draws inspiration
from the fields of signal processing and machine learning, by combining compressive sensing and Bayesian sparse learning to offer an efficient method for network reconstruction.

The paper is organised as follows. In Section \ref{sec:Model}, we introduce the type of
biochemical reaction networks model we consider in this paper, i.e., discrete Langevin ODEs. In
Section \ref{sec:Reconstruction}, we formulate the network reconstruction problem associated with the model class proposed in Section \ref{sec:Model}. In Section \ref{sec:Compressive_Sensing}, we show how the reconstruction problem can be
converted into a compressive sensing problem. In Section~\ref{sec:Bayesian}, we show how Bayesian sparse learning algorithms can be used to solve the compressive sensing problem. In
Section \ref{sec:Repressilator}, we apply our method to the reconstruction of the repressilator and show how it can be reconstructed almost exactly using the compressive sensing framework. Finally, in Section \ref{sec:Conclusion}, we conclude and discuss several future problems that we plan to address.


\section{MODEL FORMULATION}\label{sec:Model}

We consider dynamical systems of
the following general form:%
\begin{equation}
\frac{dx}{dt}=Sf(x),  \label{continuous}
\end{equation}%
where $x\in {\mathbb{R}}^{n}$, $S\in {\mathbb{R}}^{n\times L}$, and $f(\cdot
)\in {\mathbb{R}}^{L}$ is a vector of $L$ unknown functions. This set of functions
can, for example, include mass action kinetic terms under the form of product of monomials, monotonically increasing or monotonically decreasing Hill functions, simple linear terms, etc. Let neither the value of the entries nor the structure of matrix
$S$ be known. Given experimental data in terms of time series of state $%
x$, our objective is to identify the values in $S$.

Applying standard Euler discretisation to (\ref{continuous}), we obtain:
\begin{equation}
x(t_{k+1})=x(t_{k})+(t_{k+1}-t_{k})Sf(x(t_{k})),  \label{discrete-1}
\end{equation}%
Assuming that the discretisation step $t_{k+1}-t_{k}=\epsilon$ is constant for all $k$,
and defining the error between two successive state values as $e(t_{k+1}) \triangleq x(t_{k+1})-x(t_{k})$,
we can write (\ref{discrete-1}) under the general form%
\begin{equation}
e(t_{k+1})=\mathbf{v}^{T}f(x(t_{k})),  \label{discrete-2}
\end{equation}%
where $e=[e_{1},e_{2},\ldots ,e_{n}]^{T}\in {%
\mathbb{R}}^{n\times 1}$ denotes the error vector; $x=[x_{1},x_{2},\ldots ,x_{n}]^{T}\in {%
\mathbb{R}}^{n\times 1}$ denotes the state vector;
$\mathbf{v}^{T}=\left[
\begin{array}{ccc}
v_{11} & \ldots & v_{L1} \\
\vdots & \vdots & \vdots \\
v_{1n} & \ldots & v_{Ln}%
\end{array}%
\right]=\epsilon S\in {\mathbb{R}}^{n\times L}$; and $%
f(x)=[f_{1}(x),f_{2}(x),\ldots ,f_{L}(x)]^{T}\in
{\mathbb{R}}^{L\times 1}$ is the vector field.

Typically biochemical reaction networks exhibit stochastic behaviours due
to intrinsic and extrinsic noise at the gene expression level. Furthermore, the
acquisition of data is typically affected by the inaccuracy of the measurement
devices as well as by unpredictable environmental perturbations. It is therefore more realistic
to consider \textit{Langevin nonlinear dynamics}:%
\begin{equation}
e(t_{k+1})=\mathbf{v}^{T}f(x(t_{k}))+\xi (t_{k}).  \label{process-model}
\end{equation}%
$\xi (t_{k})\in {\mathbb{R}}^{n}$ represents energy-bounded process noises which are assumed to be independent, to be distributed according to normal probability distributions:%
\begin{eqnarray*}
E[\xi (t_{k})] &=&0,\text{ \ } \\
E[\xi (t_{k})\xi ^{T}(t_{j})] &=&Q_{k}\delta _{kj},
\end{eqnarray*}%
where
\begin{equation}
\delta _{kj}=\left\{
\begin{array}{ll}
1, & k=j, \\
0, & k\neq j.%
\end{array}%
\right.  \label{delta}
\end{equation}

In the following section, we shall propose an algorithm that uses
time-series data to reconstruct the eq.~\eqref{continuous}.


\section{RECONSTRUCTION PROBLEM FORMULATION}\label{sec:Reconstruction}

\subsection{Problem Formulation}

Consider the discrete-time model in eq.~\eqref{process-model}:
\begin{equation}
e^{T}(t_{k+1})=f^{T}(x(t_{k}))\textbf{v}+\xi ^{T}(t_{k}),  \label{transpose}
\end{equation}%

Assuming that $M$ successive data samples are obtained from \eqref{transpose} and defining%
\begin{eqnarray*}
\mathbf{y} &\triangleq &\left[
\begin{array}{llll}
\mathbf{y}_{1} & \ldots  & \mathbf{y}_{n}%
\end{array}\right]
=\left[ \begin{array}{llll}
e(t_1) \ldots  & e(t_M)%
\end{array}%
\right]^T  \\
&=&\left[
\begin{array}{ccc}
e_{1}(t_{1}) & \ldots  & e_{n}(t_{1}) \\
e_{1}(t_{2}) & \ldots  & e_{n}(t_{2}) \\
\vdots  & \ddots  & \vdots  \\
e_{1}(t_{M}) & \ldots  & e_{n}(t_{M})%
\end{array}%
\right] \in {\mathbb{R}}^{M\times n}, \\
\Theta  &\triangleq&\left[
\begin{array}{ccc}
f_{1}(x(t_{0})) & \ldots  & f_{L}(x(t_{0})) \\
f_{1}(x(t_{1})) & \ldots  & f_{L}(x(t_{1})) \\
\vdots  & \ddots  & \vdots  \\
f_{1}(x(t_{M-1})) & \ldots  & f_{L}(x(t_{M-1}))%
\end{array}%
\right] \\
&=&\left[
\begin{array}{c}
f^{T}(x(t_{0})) \\
f^{T}(x(t_{1})) \\
\vdots  \\
f^{T}(x(t_{M-1}))%
\end{array}%
\right] \in {\mathbb{R}}^{M\times L}, \\
\mathbf{v} &\triangleq &\left[
\begin{array}{llll}
\mathbf{v}_{1} & \mathbf{v}_{2} & \ldots  & \mathbf{v}_{n}%
\end{array}%
\right] \in {\mathbb{R}}^{L\times n},\\
\Xi  &\triangleq&\left[
\begin{array}{llll}
\xi _{1} & \xi _{2} & \ldots  & \xi _{n}%
\end{array}%
\right]  \\
&=&\left[
\begin{array}{ccc}
\xi _{1}(t_{0}) & \ldots  & \xi _{n}(t_{0}) \\
\xi _{1}(t_{1}) & \ldots  & \xi _{n}(t_{1}) \\
\vdots  & \ddots  & \vdots  \\
\xi _{1}(t_{M-1}) & \ldots  & \xi _{n}(t_{M-1})%
\end{array}%
\right] \in {\mathbb{R}}^{M\times n}.
\end{eqnarray*}

Eq.~\eqref{transpose} can be written
\begin{equation}
\mathbf{y}=\Theta \mathbf{v}+\Xi ,
\label{problem1}
\end{equation}%
or
\begin{equation}
\mathbf{y}_i=\Theta \mathbf{v}_i+\xi _i,\text{ }(i=1,\ldots ,n).
\label{problem}
\end{equation}%
We want to find $\mathbf{v}_i$ given measured data stored in $\mathbf{y}_i$.
This a typical linear regression problem that can be solved using standard least square approaches, provided that the structure of the nonlinearities in the model are known, i.e., provided that $\Theta$ is  known. However, in most cases, these nonlinearities are unknown or difficult to assume \emph{a priori}, and as a consequence, $\Theta$ itself is unknown and therefore $\mathbf{v}_i$ cannot be solved from eq.~\eqref{problem}.

Depending on the field for which the dynamical model needs to be built, only a few typical nonlinearities specific to this field need to be considered. For example, the class of models that arise from biochemical reaction networks typically involve nonlinearities that capture fundamental biochemical kinetic laws, e.g., first-order degradation functions, mass-action kinetics, Hill and Michaelis-Menten functions.
In what follows we gather in a matrix $\Phi$ similar to $\Theta$ the set of \emph{all} candidate basis functions that we want to consider for reconstruction:
\begin{equation}
\Phi \triangleq \left[
\begin{array}{ccc}
F_{1}(x(t_{0})) & \ldots  & F_{N}(x(t_{0})) \\
F_{1}(x(t_{1})) & \ldots  & F_{N}(x(t_{1})) \\
\vdots  & \ddots  & \vdots  \\
F_{1}(x(t_{M-1})) & \ldots  & F_{N}(x(t_{M-1}))
\end{array}
\right]\in\mathbb{R}^{M\times N}\label{eq:Phi}
\end{equation}
This leads to equations similar to \eqref{problem}:
\begin{equation}
\mathbf{y}_i=\Phi \mathbf{w}_i+\xi _i,\text{ }(i=1,\ldots ,n).
\label{eq:expand}
\end{equation}
where
\begin{eqnarray}
\mathbf{w}&=& \left[
\begin{array}{llll}
\mathbf{w}_{1} & \ldots  & \mathbf{w}_{n}%
\end{array}%
\right]    \label{w_omega}   \\
&=& \left[
\begin{array}{ccc}
\omega_{11} & \ldots & \omega_{1n} \\
\vdots & \vdots & \vdots \\
\omega_{N1} & \ldots & \omega_{Nn}%
\end{array}%
\right] \in {\mathbb{R}}^{N\times n}.  \notag
\end{eqnarray}
We will introduce in section \ref{sec:Compressive_Sensing} a method that allows us to reconstruct $\mathbf{w}_i$ from time-series observations of $x$ and $\mathbf{y}_i$. Before explaining this method, we introduce in the following two sections the most commonly used nonlinear functions that appear in nonlinear ODEs used to model biochemical reaction networks and explain how, based on these functions, the reconstruction problem amounts to infer $\Phi$.

\subsection{Polynomial Input Functions}\label{sec:polynomial}
A \textit{Monomial }$m_{\alpha }$ in $n$ variables is a function defined as $%
m_{\alpha }(x)=x^{\alpha }\triangleq x_{1}^{\alpha _{1}}x_{2}^{\alpha _{2}}\ldots
x_{n}^{\alpha _{n}}$ for $\alpha \in $ ${\mathbb{Z}}_{+}^{n}.$ The degree of
a monomial is defined as $\deg m_{\alpha }:=\sum\nolimits_{i=1}^{n}\alpha
_i.$ Polynomials being decomposable into sums of monomial terms, the elements $F_j(x)$ appearing in the basis function matrix $\Phi$ can be represented as monomials of the form:
\begin{equation}
F_j(x)= x_{1}^{\alpha _{1}}x_{2}^{\alpha _{2}}\ldots
x_{n}^{\alpha _{n}},  \label{eq:Fjx}
\end{equation}%
where $x_{k}$ ($k=1,\ldots ,n$) is the $k$th component of the dynamical
variable $x$.
As an illustration, we consider the case of $n=2$ variables
$x_{1}$, $x_{2}$. In that case, we have the following basis functions to represent polynomials: $F_1(x)=x_1^0x_2^0=1,~F_2(x)=x_1^1x_2^0=x_1,~F_3(x)=x_1^0x_2^1=x_2,~\ldots,~F_N(x)=x_1^{s_1}x_2^{s_2}$, where the exponents $s_1$ and $s_2$ are the maximal possible stoichiometric coefficients for $x_1$ and $x_2$ in biochemical reactions \cite{palsson2006systems}.

Given an upper bound on the degree of the polynomials that we want to consider as potential candidates in the reconstruction process, we thus need to consider as elements $F_j(x)$ of $\Phi$ all possible independent monomials of degree less or equal to the maximal degree of the candidate polynomial.


\subsection{Rational Input Functions}\label{sec:rational}
Another common type of nonlinearities in dynamical models of biochemical reaction networks are the activation and repression Hill and Michaelis-Menten functions. To include these nonlinearities as potential candidates in the reconstruction process, we consider the general multi-dimensional input function introduced in Section \ref{sec:introduction}
\begin{equation*}
G_{j}(x)=\frac{\sum\nolimits_{j}\beta
_{j}(x_{j}/K_{j})^{n_{j}}}{1+\sum\nolimits_{j}\beta _{j}(x_{j}/K_{j})^{m_{j}}%
},
\end{equation*}%
where $\beta _{j}$ and $K_{j}$ are parameters to be identified and the exponents $n_j$ and $m_j$ are assumed to be fixed. The cases of gene activation or repression are encompassed by this general form: gene activation corresponds to the situation where $n_j=m_j>0$ whereas gene repression corresponds to the case where $n_j=0$.

To give an idea on how to deal with rational rate terms, we assume a general model of the form as described in \eqref{discrete-2}:
\begin{equation}
e_i(t_{k+1})=\frac{\sum\nolimits_{j}\alpha
_{ij}(x_{ij}(t_{k})/K_{ij})^{n_{ij}}}{1+\sum\nolimits_{j}\beta
_{ij}(x_{ij}(t_{k})/K_{ij})^{m_{ij}}}-\gamma _ix_i(t_{k})+C_i.
\label{multi-input}
\end{equation}%
By multiplying both sides of equation (\ref{multi-input}) by $1+\sum\nolimits_{j}\beta _{ij}(x_{ij}(t_{k})/K_{ij})^{m_{ij}}$, we obtain%
\begin{eqnarray*}
&&\left[ 1+\sum\nolimits_{j}\beta _{ij}(x_{j}(t_{k})/K_{ij})^{m_{ij}}\right]
e_i(t_{k+1}) \\
&=&-\gamma _i\left[ 1+\sum\nolimits_{j}\beta
_{ij}(x_{j}(t_{k})/K_{ij})^{m_{ij}}\right] x_i(t_{k}) \\
&&+\left[ 1+\sum\nolimits_{j}\beta _{ij}(x_{j}(t_{k})/K_{ij})^{m_{ij}}\right] C_i
\\
&&+\sum\nolimits_{j}\alpha _{ij}(x_{j}(t_{k})/K_{ij})^{n_{ij}}
\end{eqnarray*}


\subsubsection{Activation Hill Functions: $n_{ij}=m_{ij}$}

We let
\begin{eqnarray*}
&&%
\begin{array}{rrr}
\hat{\gamma}_i=-\gamma _i, & \bar{\alpha}_{ij}=\alpha
_{ij}/K_{ij}^{n_{ij}}, & \bar{\beta}_{ij}=\beta _{ij}/K_{ij}^{n_{ij}}.%
\end{array}
\\
&&%
\begin{array}{rr}
\hat{\alpha}_{ij}=\bar{\alpha}_{ij}+C_i\bar{\beta}_{ij}, & \hat{\beta}%
_{ij}=\hat{\gamma}_i\bar{\beta}_{ij},%
\end{array}%
\end{eqnarray*}%
We then obtain:
\begin{eqnarray*}
&&\left[ 1+\sum\nolimits_{j}\bar{\beta}_{ij}x_{j}^{n_{ij}}(t_{k})\right]
e_i(t_{k+1}) \\
&=& \hat{\gamma}_i\left[ 1+\sum\nolimits_{j}\bar{\beta}%
_{ij}x_{j}^{n_{ij}}(t_{k})\right] x_i(t_{k})
+\sum\nolimits_{j}\hat{\alpha}_{ij}
x_{j}^{n_{ij}}(t_{k})+C_i
\end{eqnarray*}%
This leads to%
\begin{eqnarray*}
&&e_i(t_{k+1}) \\
&=&\hat{\gamma}_ix_i(t_{k})+\sum\nolimits_{j}\hat{\alpha}%
_{ij}x_{j}^{n_{ij}}(t_{k})+\sum\nolimits_{j}\hat{\beta}%
_{ij}x_{j}^{n_{ij}}(t_{k})x_i(t_{k}) \\
&&-\sum\nolimits_{j}\bar{\beta}%
_{ij}x_{j}^{n_{ij}}(t_{k})e_i(t_{k+1})+C_i,\\
&=&\hat{\gamma}_ix_i(t_{k})+\sum\nolimits_{j}\hat{\alpha}%
_{ij}x_{j}^{n_{ij}}(t_{k})+\sum\nolimits_{j}\hat{\beta}%
_{ij}x_{j}^{n_{ij}}(t_{k})x_i(t_{k}) \\
&&-\sum\nolimits_{j}\bar{\beta}%
_{ij}x_{j}^{n_{ij}}(t_{k})(x_i(t_{k+1})-x_i(t_{k}))+C_i\\
&=&\hat{\gamma}_ix_i(t_{k})+\sum\nolimits_{j}\hat{\alpha}%
_{ij}x_{j}^{n_{ij}}(t_{k})+\sum\nolimits_{j}\hat{\beta}%
_{ij}x_{j}^{n_{ij}}(t_{k})x_i(t_{k}) \\
&&-\sum\nolimits_{j}\bar{\beta}%
_{ij}x_{j}^{n_{ij}}(t_{k})x_i(t_{k+1})+\sum\nolimits_{j}\bar{\beta}%
_{ij}x_{j}^{n_{ij}}(t_{k})x_i(t_{k})+C_i,
\end{eqnarray*}%
from which $\hat{\gamma}_i,\hat{\alpha}_{ij},\hat{\beta}_{ij},\bar{\beta}_{ij},C_i$
can be estimated.
We can then calculate $\gamma _i,\bar{\beta}_{ij}$ and $%
\bar{\alpha}_{ij}$ .

\subsubsection{Repression Hill Functions: $n_{ij}=0$}
We let
\begin{eqnarray*}
&&%
\begin{array}{rrr}
\tilde{\gamma}_i=-\gamma _i, & \tilde{\alpha}_{ij}=\alpha _{ij}, &
\tilde{\beta}_{ij}=\beta _{ij}/K_{ij}^{m_{ij}},%
\end{array}
\\
&&%
\begin{array}{rrr}
\check{\alpha}_{ij}=C_i\tilde{\beta}_{ij}, & \check{\beta}_{ij}=\tilde{\gamma}%
_i\tilde{\beta}_{ij}, & \tilde{C}_i=\sum\nolimits_{j}\tilde{\alpha}%
_{ij}+C_i.%
\end{array}%
\end{eqnarray*}%
We then obtain:%
\begin{eqnarray*}
&&\left[ 1+\sum\nolimits_{j}\tilde{\beta}_{ij}x_{j}^{m_{ij}}(t_{k})\right]
e_i(t_{k+1}) \\
&=&\tilde{\gamma}_i\left[ 1+\sum\nolimits_{j}\tilde{\beta}%
_{ij}x_{j}^{m_{ij}}(t_{k})\right] x_i(t_{k})+\sum\nolimits_{j}C_i\tilde{%
\beta}_{ij}x_{j}^{m_{ij}}(t_{k})+\tilde{C}_i
\end{eqnarray*}%
This leads to:
\begin{eqnarray*}
&&e_i(t_{k+1}) \\
&=&\tilde{\gamma}_ix_i(t_{k})+\sum\nolimits_{j}\check{\alpha}%
_{ij}x_{j}^{m_{ij}}(t_{k})+\sum\nolimits_{j}\check{\beta}%
_{ij}x_{j}^{m_{ij}}(t_{k})x_i(t_{k}) \\
&&-\sum\nolimits_{j}\tilde{\beta}_{ij}x_{j}^{m_{ij}}(t_{k})(x_i(t_{k+1})-x_i(t_{k}))+%
\tilde{C}_i\\
&=&\tilde{\gamma}_ix_i(t_{k})+\sum\nolimits_{j}\check{\alpha}%
_{ij}x_{j}^{m_{ij}}(t_{k})+\sum\nolimits_{j}\check{\beta}%
_{ij}x_{j}^{m_{ij}}(t_{k})x_i(t_{k}) \\
&&-\sum\nolimits_{j}\tilde{\beta}_{ij}x_{j}^{m_{ij}}(t_{k})x_i(t_{k+1})+\sum\nolimits_{j}\tilde{\beta}_{ij}x_{j}^{m_{ij}}(t_{k})x_i(t_{k})+%
\tilde{C}_i
\end{eqnarray*}%
from which $\tilde{\gamma}_i,\check{\alpha}_{ij},\check{\beta}_{ij},\tilde{\beta}%
_{ij},\tilde{C}_i$ can be estimated. Based on these, we can then calculate $\gamma _i,\bar{\beta%
}_{ij}$ and $\bar{\alpha}_{ij}$.

We thus see that the problem of identification of rational polynomial functions with unknown parameters in their denominator can be reduced to a problem of polynomial identification. This can be done following the approach presented in Section \ref{sec:polynomial} by extending the basis functions.

To consider various Hill coefficients $m_{j}$, e.g., all possible Hill coefficient values from the set $\{1,2,3,4\}$, we just need to consider all the functions $G_j(x)$ with these individual values of $m_{j}$ and to add the corresponding monomial terms extracted from the above derivations as elements of the sensing matrix $\Phi$.

\subsection{Problem Transformation}
Once the matrix $\Phi$ is constructed so as to contain all the candidate basis functions that we want to consider in the reconstruction, the remaining task is to propose an efficient method that allows to solve for $\mathbf{w}_i$ in eq.~\eqref{eq:expand}:
$$\mathbf{y}_i=\Phi \mathbf{w}_i+\xi _i,\text{ }(i=1,\ldots ,n).$$


Typically the weighting vector $\mathbf{w}_i$ solution of \eqref{eq:expand} is $k$\textit{-sparse.} Mathematically, we say that a signal $\mathbf{w}$ is $k$\textit{-sparse} when it has at most \textit{k }non-zero entries, i.e., $\left\Vert \mathbf{w}\right\Vert
_{0}\leq k$. We let
\begin{equation*}
\Omega_{k}=\left\{ \mathbf{w}:\left\Vert \mathbf{w}\right\Vert
_{0}\leq k\right\}
\end{equation*}%
denote the set of all $k$\textit{-}sparse signals. On one hand, biochemical
reaction networks are typically sparse \cite{jeong2000large}. On the other
hand, the acquisition of sufficient biological time series data over long time spans is
quite difficult due to the typical cost of wet-lab experiments and current technological
limitations in terms of the type and quality of the measurements. Furthermore, since the nonlinear form of the equation is typically unknown, there can potentially be a very large number of candidate functions. As a consequence, we typically have $N\gg M$ for $\Phi_{M\times N}$ and $\mathbf{w}_i$ sparse.

The linear regression problem \eqref{eq:expand} can thus be defined as a
compressive sensing, or sparse signal recovery problem \cite%
{Candes2005decoding,donoho2006compressed},
with observation vector $\mathbf{y}_i$, known regressor matrix $\Phi,$
unknown coefficients $\mathbf{w}_i,$ and additive noise $\Xi \mathbf{.}$ In
sparse problems, the prior belief is that only a small fraction of the elements
appearing in $\mathbf{w}_i$ are non-negligible. The general aim is to identify the smallest
subset of columns of $\Phi$, whose linear span contains the observations $%
\mathbf{y}_i$. Identifying this smallest subset for all $i=1,\ldots,n$ corresponds to looking for the sparsest $\mathbf{w}$ that satisfies the following equation:
\begin{equation}\label{eq:new}
\mathbf{y}=\Phi \mathbf{w}+\Xi.
\end{equation}

The solution $\mathbf{w}^*$ to eq.~\eqref{eq:new} is, by definition, going to be sparse. Now, if all the elements in the $i^{th}$ row of $\mathbf{w}^*$ are $0$, then we can delete the $i^{th}$ row of $\mathbf{w}^*$ and the $i^{th}$ column of $\Phi$. After repeating this for all the rows of $\mathbf{w}^*$, we can equivalently rewrite \eqref{eq:new} as
\begin{equation}\label{eq:new1}
\mathbf{y}=\Phi^*\mathbf{\bar{w}}^*+\Xi^*.
\end{equation}

The solution of eq.~\eqref{eq:new1} is an estimate of the solution of the original problem defined in eq.~\eqref{problem1} which itself is related to the discretisation of eq.~\eqref{continuous}.

%



\section{COMPRESSIVE SENSING}\label{sec:Compressive_Sensing}
\subsection{Algorithm for Compressive Sensing}
To be consistent with eq.~\eqref{eq:expand}, in this and next section, $i=1,\ldots ,n$.
Since $\mathbf{w}_i$ is sparse, we use regularisation methods to recover $\mathbf{w}_i$. Regularisation methods impose upper bound constraints on the $l_{0}$ norm of $\mathbf{w}_i$, $\left\Vert \mathbf{w}_i%
\right\Vert _{0}$, i.e., on the number of nonzero terms in $\mathbf{%
w}_i$. A sparse approximation to $\mathbf{w}_i$ is readily obtained by solving the
following optimisation problem:%
\begin{equation}
\mathbf{\hat{w}_i=}\arg \min\limits_{\mathbf{w}_i}\{\left\Vert \mathbf{y}_i-\Phi
\mathbf{w}_i\right\Vert _{2}^{2}+\rho_i \left\Vert \mathbf{w}_i\right\Vert _{0}\}.
\label{L0optimisation}
\end{equation}%
Unfortunately, this optimisation problem is both numerically unstable and NP-complete;
therefore some simplifications are typically used to recast this problem into another one for which efficient algorithmic solutions exist. The most common simplification is to use the $l_{1}$-norm instead of the $l_{0}$%
-norm \cite{Candes2005decoding}, so that the optimisation problem becomes%
\begin{equation}
\mathbf{\hat{w}_i=}\arg \min\limits_{\mathbf{w}_i}\{\left\Vert \mathbf{y}_i-\Phi
\mathbf{w}_i\right\Vert _{2}^{2}+\rho_i \left\Vert \mathbf{w}_i\right\Vert _{1}\}.
\label{L1optimisation}
\end{equation}
The approach presented in eq.~\eqref{L1optimisation} is known as \emph{Lasso} \cite%
{tibshirani1996regression}. A number of methods have been proposed to solve Compressive Sensing (CS) reconstruction problems, including $l_{1}$%
-minimisations \cite{chen1999atomic,Candes2005decoding}
and greedy algorithms \cite%
{donoho2005sparse,tropp2007signal,dai2009subspace}. Both methods can actually be shown to have performance guarantees for the exact recovery of $\mathbf{w}_i$ if $\Phi$ satisfies the so called \textit{restricted isometry property} \cite{Candes2005decoding}. We will discuss this in more detail in Section \ref{sec:Conclusion}.

Since $\Phi$ as defined in (\ref{eq:Phi})
could be rank deficient, \eqref{L1optimisation} cannot, in general, be
readily solved by these two algorithms
and thus a different approach needs to be considered. Thanks to recent results in signal processing and machine learning, we propose hereafter a method to solve the compressive sensing problem defined in \eqref{L1optimisation} using a
Bayesian formulation \cite{tipping2001sparse,tipping2002analysis,tipping2003fast,ji2008bayesian}.


\subsection{Bayesian Compressive Sensing}


\subsubsection{Linear Regression and Maximum Likelihood}
Bayesian modelling treats all unknowns as
stochastic variables with certain probability distributions \cite{bishop2006pattern}.
Using a Bayesian framework, we assume that the stochastic variable $\xi_i$ are independent and characterised by a Gaussian distribution with zero-mean and variance $\sigma_i^2$.  We further define the precision or inverse-variance as $\beta_i =1/\sigma_i^{2}$.

The multivariate Gaussian likelihood of $\mathbf{y}_i$ is then given by:
\begin{eqnarray}
p(\mathbf{y}_i|{\mathbf{w}_i},\beta_i)
&=&{\mathcal{N}}(\mathbf{y}_i|\Phi {\mathbf{w}_i},\beta_i^{-1})  \notag \\
&=&(2\pi \sigma_i^{2})^{-\frac{M}{2}}\exp \left( -\frac{1}{2\sigma_i^{2}}%
\left\Vert \mathbf{y}_i-\Phi {\mathbf{w}_i}\right\Vert ^{2}\right) \notag\\
\label{likelihood}
\end{eqnarray}%
Obtaining maximum likelihood estimates for $\mathbf{w}_i$ under these
conditions is equivalent to searching for a minimal $l_{2}$-norm solution to eq.~\eqref{problem1}.

\subsubsection{Sparseness Prior}
The sparseness of $\mathbf{w}_i$ can be imposed by using a
Laplace prior of the form \begin{equation}
p({\mathbf{w}_i}|\lambda )=\frac{\lambda_i}{2}\exp (-\lambda_i\left\Vert \mathbf{%
w}_i\right\Vert _{1}).  \label{laplace prior}
\end{equation}
This then becomes equivalent to the $l_{1}$-regularisation
formulation in eq.~\eqref{L1optimisation}.
Using the likelihood function in eq.~\eqref{likelihood} and the prior in eq.~\eqref{laplace prior}, the solution of eq.~\eqref{L1optimisation} can be shown to correspond to a \textit{maximum a posteriori} (MAP) estimate for~$\mathbf{w}_i$.

However, obtaining an estimate of the full
posteriors on $\mathbf{w}_i$ and $\beta_i$ cannot be accomplished using the Laplace prior directly since Laplace priors are not conjugate to the conditional
Gaussian distribution in eq.~\eqref{likelihood}.
In the following section, we show that the use of hierarchical priors alleviates this. The CS problem can thus be converted into a linear-regression problem with a prior which is sparse.
Given the sensing matrix $\Phi$, we show in the next section how the sparse weights $\mathbf{w}_i$
and the inverse of noise variance $\beta_i$ can be estimated using in a sparse Bayesian learning approach.

\section{RECONSTRUCTION VIA BAYESIAN COMPRESSIVE SENSING}\label{sec:Bayesian}

\subsection{Specification of Hierarchical Priors}
Instead of imposing a Laplace prior on $\mathbf{w}_i$, as defined in the last section, a sparse Bayesian learning approach is adopted \cite{tipping2001sparse}. Namely, we use hierarchical priors over the distribution of $\mathbf{w}_i$. This allows for a convenient conjugate-exponential analysis with
similar properties as for the Laplace priors. The main advantage of such an approach is that it allows us to impose a Gaussian
prior with zero-mean on each element of $\mathbf{w}_i$ (see \eqref{w_omega}), i.e.,
\begin{equation}
p({\mathbf{w}_i}|\boldsymbol{\alpha}_i)=\prod\limits_{j=1}^{N}{\mathcal{N}}(\omega
_{ij}\mathbf{|}0,\alpha _{ij}^{-1}).  \label{gaussian prior}
\end{equation}%
In \eqref{gaussian prior}, $\boldsymbol{\alpha}_i=(\alpha _{i1},\ldots,\alpha _{iN})\in\mathbb{R}^{1\times N}$ represent a vector of
 $N$ independent hyperparameters,
with $\alpha _{ij}$ controlling the precision (or the inverse of the variance) of the prior imposed on $\mathbf{w}_i$.
It is this form of prior that is eventually responsible for the sparsity properties of the model (see \cite{tipping2001sparse} for more details).
It is common to place a Gamma prior
on $\boldsymbol{\alpha}_i$ as follows:
\begin{equation}
p(\boldsymbol{\alpha}_i|a,b)=\prod\limits_{j=1}^{N}\Gamma (\alpha_{ij}\mathbf{|}a,b).  \label{gamma prior}
\end{equation}%
where the Gamma distribution is defined as:%
\begin{equation*}
\text{$\Gamma $}(\xi \mathbf{|}a,b)=\frac{\left( b\right) ^{a}}{\Gamma (a)}%
\xi ^{a-1}\exp [-b\xi ]
\end{equation*}%
where $\Gamma (a)=\int_{0}^{\infty }t^{a-1}e^{-t}dt$ is called the `Gamma function',
$\xi >0$ denotes a hyperparameter, and $%
a>0$ is the shape parameter, $b>0$ is a scaling parameter.
The $\Gamma $ distribution is generally chosen as the prior
for the precision of a Gaussian distribution because (a) it corresponds to its conjugate prior, thereby greatly simplifying the analysis and (b) it also includes the uniform
distribution as a limiting case. The overall prior on $\mathbf{w}_i$ is then
evaluated as
\begin{equation}
p({\mathbf{w}_i}\mathbf{|}a,b)= \prod\limits_{j=1}^{N}\int\nolimits_{0}^{\infty
}{\mathcal{N}}(\omega _{ij}\mathbf{|}0,\alpha _{ij}^{-1})\text{$\Gamma $}%
(\alpha _{ij}\mathbf{|}a,b)d\alpha _{ij}.  \label{overall prior}
\end{equation}%
The density function $\Gamma (\alpha_{ij}\mathbf{|}a,b)$ is the conjugate
prior for $\alpha_{ij}$ when $\omega_{ij}$ plays the role of observed data
and ${\mathcal{N}}(\omega_{ij}\mathbf{|}0,\alpha_{ij}^{-1})$ is the associated likelihood
function. Based on this, the integral $\int\nolimits_{0}^{\infty }{\mathcal{N}%
}(\omega_{ij}\mathbf{|}0,\alpha_{ij}^{-1})\Gamma (\alpha_{ij} |%
a,b)d\alpha_{ij}$ can be evaluated analytically. It can be shown that this integral
corresponds to the Student's \emph{t}-distribution which is strongly peaked around $\omega_{ij}=0$.
Consequently, the prior in (\ref{overall prior}) is a sparseness prior for $\mathbf{w}_i$.
Similarly, a Gamma prior is introduced on $\beta_i$
\begin{equation}
p(\beta_i|c,d) = \Gamma(\beta_i\mathbf{|}c,d). \label{gamma prior2}
\end{equation}%


\subsection{Bayesian Inference via Relevance Vector Machine}

Given $\mathbf{y}_i$ and $\mathbf{\Phi}$ and assuming that the hyperparameters $\boldsymbol{\alpha}_i$ and $\beta_i$ are known, the \emph{posterior} distribution for ${\mathbf{w}_i}$ conditioned on the data is given by combining the likelihood and prior with Bayes' rule:
\begin{equation}
p(\mathbf{w}_i|\mathbf{y}_i,\boldsymbol{\alpha}_i,\beta_i)
=\frac{p(\mathbf{y}_i|\mathbf{w}_i,\beta_i)p(\mathbf{w}_i|\boldsymbol{\alpha}_i)}
{p(\mathbf{y}_i|\boldsymbol{\alpha}_i, \beta_i)}. \label{Bayes}
\end{equation}%
This posterior distribution of $\mathbf{w}_i$ is Gaussian and the associated
mean and covariance matrices are given as%
\begin{eqnarray}
\mathbf{m}_i &=&\beta_i \mathbf{\Sigma \Phi}^{T}\mathbf{y}_i  \label{mean} \\
\mathbf{\Sigma}_i &=&(\mathbf{A}_i+\beta_i \mathbf{\Phi}^{T}\mathbf{\Phi })^{-1}
\label{covariance}
\end{eqnarray}%
where $\mathbf{A}_i=\mathrm{diag}(\alpha_{ij}).$ In the context of Relevance Vector Machines (RVM),
the associated ``learning" problem becomes the search for the hyperparameters
$\boldsymbol{\alpha}_i$ and $\beta_i$. RVM is a Bayesian sparse kernel technique
for regression and classification \cite{tipping2001sparse}.
It shares a lot of characteristics with support vector machines \cite{bishop2006pattern}.
In RVM, these hyperparameters are
estimated from the data by maximising
\begin{equation*}
p(\mathbf{y}_i|\boldsymbol{\alpha}_i,\beta_i )=\int p(\mathbf{y}_i|w_i,\beta_i )p(\mathbf{w}_i|\boldsymbol{\alpha}_i%
)d\mathbf{w}_i.
\end{equation*}%
This quantity is known as the marginal likelihood. The maximisation is known as a type-2 maximum likelihood \cite{berger1985statistical} or evidence approximation maximisation \cite{mackay1992bayesian}.

To avoid the convolution of two Gaussians, one can
use log marginal likelihood ${\mathcal{L}}\left( \boldsymbol{\alpha}_i,\beta_i \right)$. This leads to
\begin{eqnarray}
{\mathcal{L}}\left( \boldsymbol{\alpha}_i,\beta_i \right) &=&\log p(\mathbf{%
y}_i|\boldsymbol{\alpha}_i,\beta_i ) \notag \\
&=&\log \int p(\mathbf{y}_i|w_i,\beta_i)p(\mathbf{w}_i|\boldsymbol{\alpha}_i)d\mathbf{w}_i \notag \\
&=&\log {\mathcal{N}}(\mathbf{y}_i|0,\mathbf{C}_i) \notag \\
&=&-\frac{1}{2}\left[ M\log 2\pi +\log \left\vert \mathbf{C}_i\right\vert +%
\mathbf{y}_{i}^{T}\mathbf{C}_{i}^{-1}\mathbf{y}_{i}\right]  \notag \\
\label{logarithm}
\end{eqnarray}%
where the $M\times M$
matrix $\mathbf{C}_i=\sigma_i^{2}\mathbf{I+\Phi A}^{-1}_i\mathbf{\Phi }^{T}.$
A type-2 maximum likelihood approximation employs the point estimates for $%
\boldsymbol{\alpha}_i$ and $\beta_j$ to maximise ${\mathcal{L}}\left( \boldsymbol{\alpha}_i,\beta_j \right)$.
By setting the required derivatives of the marginal likelihood to zero, we can obtain
the following re-estimate equations \cite{mackay1992bayesian,tipping2001sparse}
\begin{equation}
\alpha_{ij}^{new}=\frac{\gamma_{ij}}{m_{ij}^{2}}, (j=1,2,\ldots
,N),  \label{update1}
\end{equation}%
where $\mathbf{m}_i$ is the $i$th posterior mean weight appearing in (\ref{mean}) and $%
\gamma_{ij}\triangleq 1-\alpha_{ij}\mathbf{\Sigma}_{i}(jj)$, with $\mathbf{\Sigma}_{i}(jj)$ representing the $j$th diagonal element of the posterior weight covariance
defined in (\ref{covariance}). Estimation of $\gamma_{ij}$ and $m_{ij}$ can be efficiently done via the expectation-maximisation algorithm, an iterative procedure for maximum likelihood parameter estimation from data sets with missing or hidden variables \cite{dempster1977maximum}.

For a noise variance $\sigma^{2}_i=1/\beta_i$, the re-estimate equation is%
\begin{equation}
\left( \beta_i^{new}\right)^{-1}=\frac{\left\Vert \mathbf{y}_i-\mathbf{\Phi m}_i%
\right\Vert _{2}^{2}}{M-\sum\limits_j\gamma _{ij}}.  \label{update2}
\end{equation}%
The quantity$\ \gamma_{ij}$ in \eqref{update2} measures how
well the corresponding parameter $\omega_{ij}$ appearing in \eqref{eq:expand} is determined by the data.
Note that $\alpha_{ij}^{new}$, $j=1,\dots,N$ and $\beta_i^{new}$ are functions of $%
\mathbf{m}_i$ and $\mathbf{\Sigma}_i$ in equations \eqref{update1}-\eqref{update2}. Furthermore, $\mathbf{m}_i$ and $\mathbf{\Sigma}_i$
are a function of $\boldsymbol{\alpha}_i$ and $\beta_i$ in equations \eqref{mean}-\eqref{covariance}. This suggests an
iterative algorithm, which iterates between equations \eqref{mean}-\eqref{covariance}
and equations \eqref{update1}-\eqref{update2} until a convergence criterion has been satisfied.
During re-estimation, many of the $\alpha_{ij}$
tend to infinity when the corresponding $\omega_{ij}$ are very small or zero.
Usually, only a few $\alpha_{ij}$ are small.
From \eqref{Bayes}, this implies that $p(\omega_{ij}|\mathbf{y}_i,\boldsymbol{\alpha}_i,\beta_i)$ becomes highly (in principle, infinitely) peaked at zero. This implies that we can be \emph{a posteriori} `certain' that $\omega_{ij}$ are zero. The corresponding basis functions in $\Phi$ can thus be `pruned', and sparsity is realised \cite{tipping2001sparse}.
In \cite{tipping2002analysis} and \cite{tipping2003fast}, a fast RVM algorithm was
developed by analysing the properties of the log marginal likelihood function in
eq.~\eqref{logarithm}. This enables a principled and efficient sequential
addition and deletion of candidate basis function (columns of $\mathbf{\Phi }
$) to monotonically maximise the log marginal likelihood. Details about this fast
algorithm can be found in \cite{tipping2002analysis} and \cite{tipping2003fast}.


\section{IDENTIFICATION OF THE REPRESSILATOR}\label{sec:Repressilator}


We consider here a classical dynamical system in systems/synthetic biology which
we will use to illustrate the biochemical network reconstruction problem at hand.
The repressilator is a synthetic oscillator network that was
originally conceived and constructed by Elowitz and Leibler \cite%
{elowitz2000son}. The network consists of three genes repressing each other in a ring structure.

\begin{figure}[h]
\center
\includegraphics[width=9cm]{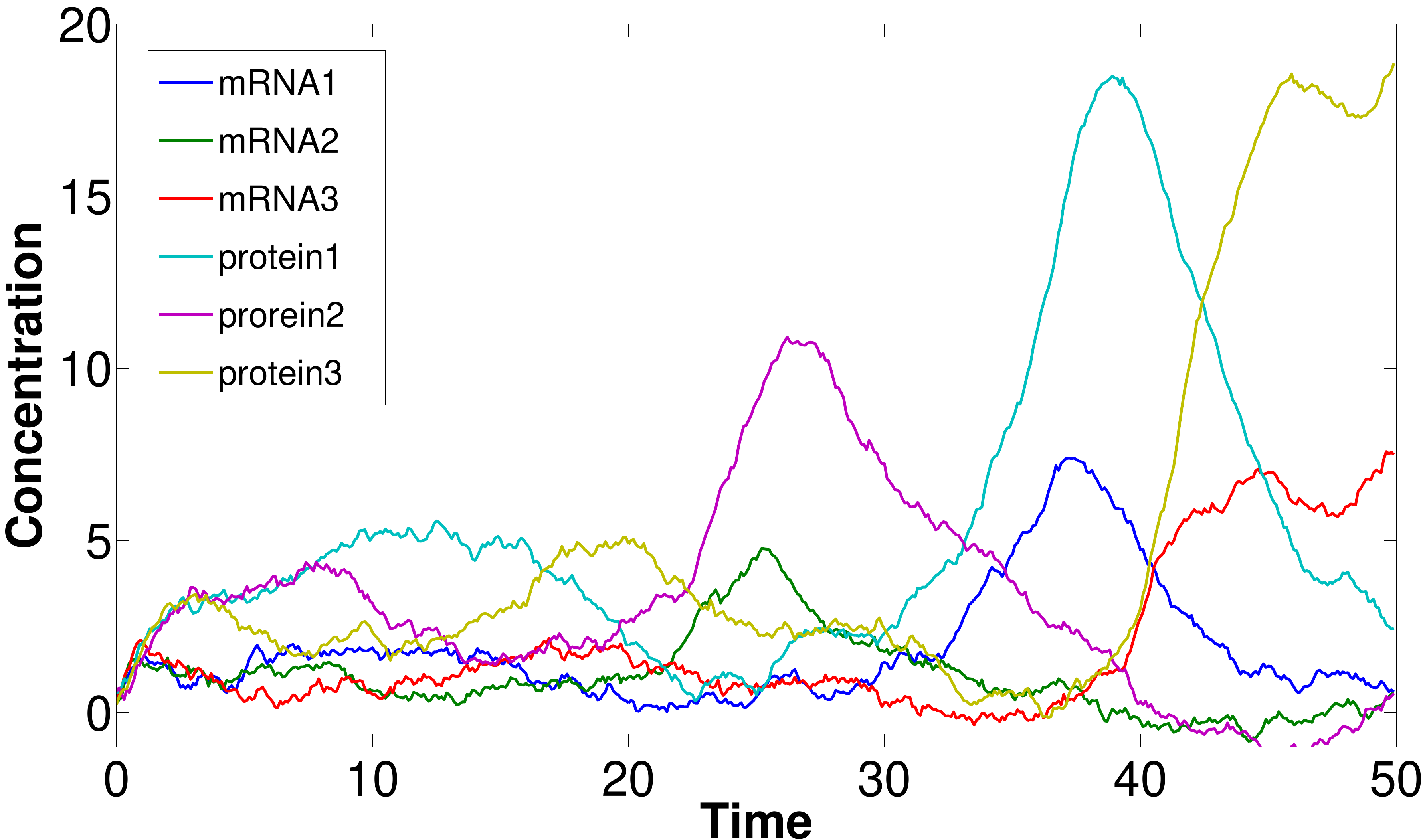}
\caption{Time series of $x_{1},\dots,x_{6}$ for the repressilator model (\protect
\ref{oscillator}).}\label{fig:Euler_simulations}
\end{figure}

A mathematical description the repressilator that includes both
transcription and translation dynamics can be described as
\begin{eqnarray}
\frac{dx_{1}}{dt} &=&-\gamma _{1}x_{1}+\frac{\alpha _{1}}{1+x_{6}^{n_{1}}}%
+\theta _{1},  \notag \\
\frac{dx_{2}}{dt} &=&-\gamma _{2}x_{2}+\frac{\alpha _{2}}{1+x_{4}^{n_{2}}}%
+\theta _{2},  \notag \\
\frac{dx_{3}}{dt} &=&-\gamma _{3}x_{3}+\frac{\alpha _{3}}{1+x_{5}^{n_{3}}}%
+\theta _{3},  \notag \\
\frac{dx_{4}}{dt} &=&-\gamma _{4}x_{4}+\beta _{1}x_{1},  \notag \\
\frac{dx_{5}}{dt} &=&-\gamma _{5}x_{5}+\beta _{2}x_{2},  \notag \\
\frac{dx_{6}}{dt} &=&-\gamma _{6}x_{6}+\beta _{3}x_{3}, \label{oscillator}
\end{eqnarray}%
Here, $x_{1},x_{2},x_{3}$ denote the concentrations of the mRNA
transcripts of genes 1, 2, and 3, respectively whereas
$x_{4},x_{5},x_{6}$ denote the protein concentrations of the respective
genes.
$\alpha _{1},\alpha
_{2},\alpha _{3}$ denote the maximum promoter strength for their corresponding gene,
$\gamma_{1},\gamma _{2},\gamma _{3}$ denote the mRNA decay rate, $\gamma
_{4},\gamma _{5},\gamma _{6}$ denote the protein decay rate, $\beta
_{1},\beta _{2},\beta _{3}$ denote the protein production rate, $\theta
_{1},\theta _{2},\theta _{3}$ denote the basal transcription rate. The set of ODEs in (\ref%
{oscillator}) corresponds to a topology where gene $1$ is repressed by gene $2$,
gene $2$ is repressed by gene $3$, and gene $3$ is repressed by gene $1$. Using the
standard forward Euler method to numerically solve ODEs, we obtain trajectory of the
six states $x_{1},\dots,x_{6}$ in Fig.~\ref{fig:Euler_simulations}. These trajectories are then sampled
to generate a time-series of gene expression data.

Take gene 1 for example. The hill coefficient $n_{1}$ will typically have a value within a range from $1$ to $4$ due to biochemical constraints. The core question here is:
how can we determine the topology and kinetic parameters of the set of ODEs in (\ref{oscillator}%
)? Note that we do not assume \emph{a priori} knowledge of the form of the nonlinear functions, e.g., whether the degradation obeys first-order or enzymatic catalysed dynamics or whether the proteins are repressors or activators.


Next we show how the network construction problem of the repressilator model in \eqref{oscillator} can be formulated under the form presented in \eqref{eq:expand}. 

Following the procedure in \eqref{discrete-1}, \eqref{discrete-2} and \eqref{transpose}, we construct a matrix of candidate functions $\Phi$ by selecting the most commonly used candidate basis functions used to model biochemical reaction networks (see Sections \ref{sec:polynomial} and \ref{sec:rational}).
As a proof of concept, we only consider Hill functions as potential nonlinear candidates. The set of Hill functions with Hill coefficient $i$, both in activating and repressing from, for each of the $6$ state variables reads:
\begin{eqnarray*}
\text{hill}_i(t_{k}) &\triangleq &\left[ \frac{1}{1+x_{1}^{i}(t_{k})},\ldots ,%
\frac{1}{1+x_{6}^{i}(t_{k})},\right.  \notag \\
&&\left. \frac{x_{1}^{i}(t_{k})}{1+x_{1}^{i}(t_{k})},\ldots ,\frac{%
x_{6}^{i}(t_{k})}{1+x_{6}^{i}(t_{k})}\right]_{1\times 12}, 
\end{eqnarray*}%
where $i$ represents the Hill coefficient. In what follows we consider
that the Hill coefficient can take any of the following values: $1$, $2$, $3$ or $4.$ To also take into account basal transcription expression rates, we add to the last column of $\Phi$ a unit vector. Since there are 6 state variables, we can construct the basis matrix $\Phi$ appearing in (\ref{oscillator_phi}) with $6$ (basis functions for linear terms) $+(4*12)=48$ (basis functions for hill functions) $+1$ (basis function for basal expression) $=55$ columns.

\begin{figure*}[t]
\begin{equation}
\Phi =\left[
\begin{array}{ccccccc}
x_{1}(t_{1}) & \ldots & x_{6}(t_{1}) & \text{hill}_{1}(t_{1}) & \ldots &
\text{hill}_{4}(t_{1}) & 1 \\
\vdots &  & \vdots & \vdots &  & \vdots & \vdots \\
x_{1}(t_{K-1}) & \ldots & x_{6}(t_{K-1}) & \text{hill}_{1}(t_{K-1}) & \ldots &
\text{hill}_{4}(t_{K-1}) & 1%
\end{array}%
\right] _{(K-1)\times (6+48+1)}  \label{oscillator_phi}
\end{equation}%
\end{figure*}

Considering \eqref{eq:expand} with the basis function matrix $\Phi$ given in \eqref{oscillator_phi}, the corresponding target weight matrix $\mathbf{w}$ should be:%
\begin{equation*}
{\small
\left[
\begin{array}{cccccc}
-\gamma _{1} & 0 & 0 & \beta _{1} & 0 & 0 \\
0 & -\gamma _{2} & 0 & 0 & \beta _{2} & 0 \\
0 & 0 & -\gamma _{3} & 0 & 0 & \beta _{3} \\
0 & 0 & 0 & -\gamma _{4} & 0 & 0 \\
0 & 0 & 0 & 0 & -\gamma _{5} & 0 \\
0 & 0 & 0 & 0 & 0 & -\gamma _{6} \\
0_{17\times 1} & 0_{15\times 1} & 0_{16\times 1} &  &  &  \\
\alpha _{1} & \alpha _{2} & \alpha _{3} & 0_{48\times 1} & 0_{48\times 1} &
0_{48\times 1} \\
0_{30\times 1} & 0_{32\times 1} & 0_{31\times 1} &  &  &  \\
\theta _{1} & \theta _{2} & \theta _{3} & 0 & 0 & 0%
\end{array}%
\right] .
}
\end{equation*}%
As an illustration, we choose the parameter values for $\mathbf{w}$ as:%
\begin{equation*}
{\small
\left[
\begin{array}{cccccc}
-0.3 & 0 & 0 & 1.4 & 0 & 0 \\
0 & -0.4 & 0 & 0 & 1.5 & 0 \\
0 & 0 & -0.5 & 0 & 0 & 1.6 \\
0 & 0 & 0 & -0.2 & 0 & 0 \\
0 & 0 & 0 & 0 & -0.4 & 0 \\
0 & 0 & 0 & 0 & 0 & -0.6 \\
0_{17\times 1} & 0_{15\times 1} & 0_{16\times 1} &  &  &  \\
4 & 3 & 5 & 0_{48\times 1} & 0_{48\times 1} & 0_{48\times 1} \\
0_{30\times 1} & 0_{32\times 1} & 0_{31\times 1} &  &  &  \\
0.02 & 0.02 & 0.01 & 0 & 0 & 0%
\end{array}%
\right] .
}
\end{equation*}%
The algorithm is implemented in MATLAB R2009a. The calculation is run on a
standard laptop computer (Intel Core2 Duo P8600 2.4GHz with 4GB RAM). We run
the algorithm for the following two experiments independently with 100
rounds respectively. The sparsest and closest solutions were selected.
The computation time for each round is less than \textbf{0.1} second.

1) Let $t_{T}=50,$ sampling interval $t_{k+1}-t_{k}=0.1,$ and $T=500$. The variance of noise is fixed at $10^{-3}$. The
estimated $\mathbf{w}$ is
\begin{equation*}
{\small
\left[
\begin{array}{cccccc}
0.300 & 0 & 0 & 1.401 & 0 & 0 \\
0 & -0.398 & 0 & 0 & 1.502 & 0 \\
0 & 0 & -0.499 & 0 & 0 & 1.600 \\
0 & 0 & 0 & -0.200 & 0 & 0 \\
0 & 0 & 0 & 0 & -0.400 & 0 \\
0 & 0 & 0 & 0 & 0 & -0.600 \\
0_{17\times 1} & 0_{15\times 1} & 0_{16\times 1} &  &  &  \\
3.999 & 3.002 & 5.001 & 0_{48\times 1} & 0_{48\times 1} & 0_{48\times 1} \\
0_{30\times 1} & 0_{32\times 1} & 0_{31\times 1} &  &  &  \\
0.019 & 0.019 & 0.0093 & 0 & 0 & 0%
\end{array}%
\right].
}
\end{equation*}%
\qquad

2) Let $t_{T}=5,$ sampling interval $t_{k+1}-t_{k}=0.1,$ and $T=50$. The variance of
noise is fixed at $10^{-3}$. The estimated $\mathbf{w}$ is
\begin{equation*}
{\small
\left[
\begin{array}{cccccc}
-0.296 & 0 & 0 & 1.389 & 0 & 0 \\
0 & -0.395 & 0 & 0 & 1.511 & 0 \\
0 & 0 & -0.500 & 0 & 0 & 1.607 \\
0 & 0 & 0 & -0.197 & 0 & 0 \\
0 & 0 & 0 & 0 & -0.406 & 0 \\
0 & 0 & 0 & 0 & 0 & -0.601 \\
0_{17\times 1} & 0_{15\times 1} & 0_{16\times 1} &  &  &  \\
4.066 & 3.033 & 5.028 & 0_{48\times 1} & 0_{48\times 1} & 0_{48\times 1} \\
0_{30\times 1} & 0_{32\times 1} & 0_{31\times 1} &  &  &  \\
0 & 0 & 0 & 0 & 0 & 0%
\end{array}%
\right] ,
}
\end{equation*}%
where the basal rate could not be estimated in the lesser sample case.

\section{CONCLUSION AND DISCUSSION}\label{sec:Conclusion}
In this paper, a new network reconstruction method for biochemical reaction
networks is proposed. This method takes advantage of compressive sensing and
sparse Bayesian learning. The proposed method only requires time series data and does not assume prior knowledge about the
model structure (topology and nonlinear functional forms) and parameters.
The problem is posed in such a way that candidate nonlinear functions specific to the type of models used (here biochemical reaction networks) are sought after.
Since biochemical reaction networks are typically sparse, the key idea is to adopt a formulation which allows to transform the nonlinear identification problem into a compressive sensing problem and to solve it efficiently using a sparse Bayesian learning approach.
We have illustrated how this approach can be used to efficiently reconstruct the nonlinear ODEs of a repressilator based on time series data.

However, beyond the results presented here, some issues remain to be further
discussed and are part of ongoing work in our group.

\subsection{Performance Guarantee}
A sufficient condition for exact reconstruction with either $l_{1}$-minimisation or greedy algorithms
is the so called \textit{restricted
isometry property} (RIP): A matrix $\Phi \in {%
\mathbb{R}}^{M\times N}$
is said to satisfy the RIP with coefficients $(K,\delta )$ for $K\leq M$, $0\leq \delta \leq 1$, if for all index sets $I\subset \{1,\ldots ,N\}$ such that $\left\vert
I\right\vert \leq K$ and for all $q\in {\mathbb{R}}^{\left\vert I\right\vert
},$ one has%
\begin{equation*}
(1-\delta )\left\Vert q\right\Vert _{2}^{2}\leq \left\Vert \Phi
_iq\right\Vert _{2}^{2}\leq (1+\delta )\left\Vert q\right\Vert _{2}^{2}
\end{equation*}%
where $\Phi_i$ denotes the matrix formed by the columns of $\Phi $ with
indices in $I$. It was shown in \cite%
{Candes2005decoding,dai2009subspace,Candes2006stable} that both $l_{1}$%
-minimisations and greedy algorithms lead to exact reconstruction of $K$-sparse
signals if the matrix $\Phi $ satisfies the RIP with a constant parameter $0\leq \delta \leq 1$.

However, the Bayesian method proposed in this paper has no known performance guarantee in terms of exact recovery equivalent to the RIP condition. Establishment of such performance guarantees in the Bayesian compressive sensing framework is part of ongoing work in our group.

\subsection{Hidden Nodes}
We have so far assumed that the system is fully observable. However,
in reality, measurement observations will typically be partial \cite{yuan2011robust} (in particular, the number of hidden/unobservable nodes and their position in the network is usually unknown).
How to generalise our framework to the case of hidden nodes and partial observations is another topic of further study in our group.


\section{ACKNOWLEDGEMENT}

The authors gratefully acknowledge the support of Microsoft Research through
the PhD Scholarship Programs of Mr Wei Pan and Dr Ye Yuan. Dr Ye Yuan
acknowledges the support from EPSRC (project EP/I03210X/1). Dr Guy-Bart
Stan gratefully acknowledges the support of the EPSRC Centre for Synthetic
Biology and Innovation at Imperial College, London. The authors would like
to thank Dr Wei Dai (at Imperial College London) for the constructive
discussions that lead to this paper.

\linespread{.99}{

\end{document}